# Fuel-Economical Distributed Model Predictive Control for Heavy-Duty Truck Platoon

Mehmet Fatih Ozkan, *Graduate Student Member, IEEE* and Yao Ma, *Member, IEEE*

*Abstract*— This paper proposes a fuel-economical distributed model predictive control design (Eco-DMPC) for a homogenous heavy-duty truck platoon. The proposed control strategy integrates a fuel-optimal control strategy for the leader truck with a distributed formation control for the following trucks in the heavy-duty truck platoon. The fuel-optimal control strategy is implemented by a nonlinear model predictive control (NMPC) design with an instantaneous fuel consumption model. The proposed fuel-optimal control strategy utilizes the preview information of the preceding traffic to achieve the fuel-economical speed planning by avoiding energy-inefficient maneuvers, particularly under transient traffic conditions. The distributed formation control is designed with a serial distributed model predictive control (DMPC) strategy with guaranteed local and string stability. In the DMPC strategy, each following truck acquires the future predicted state information of its predecessor through vehicle connectivity and then applies local optimal control to maintain constant spacing. Simulation studies are conducted to investigate the fuel economy performance of the proposed control strategy and to validate the local and string stability of the platoon under a realistic traffic scenario. Compared with a human-operated platoon and a benchmark formation-controlled platoon, the proposed Eco-DMPC significantly improves fuel economy and road utilization.

I. INTRODUCTION

Road freight transportation growth leads to an increased number of heavy-duty trucks on roads and, consequently, CO2 emissions and fuel consumption significantly increase due to heavy traffic. According to a study in [1], road transportation is responsible for roughly 27% of the energy consumption of the European Union. Moreover, 20% of the responsible carbon emissions of the vehicles come from heavy-duty trucks, as stated in [2]. The recent advancements of intelligent transportation systems (ITS) have provided solutions to improve the fuel economy of heavy-duty trucks and carbon emissions with automated platooning systems [3]. Specifically, the heavy-duty trucks cruising in a platoon with small inter-platoon distances can achieve reduced air drag force acting on the trucks, thus the fuel economy of the platoon can be improved [4].

With the growing technology in Advanced Driver Assistance Systems (ADAS) and vehicle connectivity, controlling an automated heavy-duty truck platoon in a form is feasible with different control strategies. The main objective of the commonly used control strategies for the truck platoon is to maintain constant inter-platoon gap distances [5-7] or constant time headway between the trucks [8-9] with some stability properties [10-11] during steady traffic flow. However, in transient traffic conditions, these spacing control strategies lead to excessive braking and acceleration maneuvers to maintain the desired inter-platoon gap distances during the trip. Since the frequent acceleration and braking maneuvers increase the fuel consumption of the vehicles unnecessarily, these strategies are unsuitable for realistic transient traffic conditions to achieve significant fuel economy improvement for the truck platoon, even though the small inter-platoon gap distances provide reduced air drag force. Therefore, it is necessary to establish a control strategy to achieve the fuel-economical trip for the truck platoon by avoiding energy-inefficient maneuvers under transient traffic conditions. By utilizing the preview traffic information through vehicle connectivity [12] or driver behavior prediction [13], autonomous vehicles can optimally plan their operations to achieve fuel-optimal speed profiles. This fuel-optimal speed strategy is commonly referred to as eco-driving in the literature [12-19]. In [12], an eco-driving strategy for an automated vehicle is proposed utilizing the anticipatory speed information from the preceding traffic through V2V communication. Simulations results show that more than 40% of fuel economy improvement can be achieved with the proposed eco-driving strategy in real traffic scenarios, particularly under transient traffic conditions.

This study aims to investigate the potential benefits of integrating eco-driving and platooning strategies to achieve fuel-economical trips while ensuring local and string stability for the heavy-duty truck platoon in realistic and transient traffic scenarios. The distinct contributions of this study include: 1) a fuel-efficient longitudinal control design for the leader truck is proposed to improve fuel economy by avoiding energy-inefficient maneuvers. 2) the distributed control design for the following trucks is implemented with sufficient conditions to ensure local and string stability. Moreover, the performance of the proposed control strategy for the truck platoon is qualitatively analyzed and compared with a human-driven platoon and a benchmark formation-controlled platoon in the simulation studies.

The remainder of this paper is organized as follows. In Section II, the vehicle dynamics and fuel consumption models are developed. In Section III, the fuel-economical distributed control design and stability analysis are derived. In Section IV, the simulation results along with the performance analysis are presented. Section V concludes the paper.

M. F. Ozkan and Y. Ma are with the Department of Mechanical Engineering, Texas Tech University (e-mail: mehmet.ozkan@ttu.edu and yao.ma@ttu.edu).

## II. MODELING APPROACH

A homogenous heavy-duty truck platoon with a predecessor-following (PF) communication topology is considered in this study. The platoon consists of one leader truck $(i=0)$ which follows a human-driven vehicle (HDV) in traffic and three following trucks $(i=1,2,3)$, as shown in Fig. 1. In the remainder of this section, the longitudinal vehicle dynamics of the platoon and fuel consumption model will be presented.

### A. Vehicle Dynamics Model

In this study, two different control objectives are considered for the longitudinal control of the truck platoon. The first control objective is for the leader truck to achieve a fuel-optimal driving strategy by avoiding unnecessary braking and acceleration maneuvers. The second control objective is for the following trucks to regulate the inter-platoon distance gap with a constant distance policy and to achieve zero speed difference to their predecessors in the platoon.

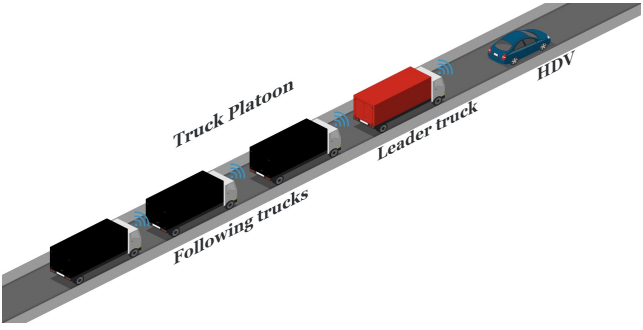

Fig. 1  Schematic of the truck platoon in traffic.

At time $t$, the individual vehicle dynamics in the platoon can be represented by the nonlinear third-order model [20] as

$$\dot{d}_i(t) = v_{i-1}(t) - v_i(t)$$
$$\dot{v}_i(t) = a_i(t) \quad (1)$$
$$\dot{a}_i(t) = f_i(v_i, a_i) + g_i(v_i)\varepsilon_i$$

where $i$ represents the $i$th vehicle in the platoon, $d_i$ represents the inter-vehicle distance of a pair of predecessor-follower vehicles, $v_i$ and $a_i$ represent longitudinal velocity and acceleration, respectively, $\varepsilon_i$ represents engine input, $f_i(.,.)$ and $g_i(.)$ are given as

$$f_i(v_i, a_i) = -\frac{1}{\tau_i}\left(a_i + \frac{\sigma A_{F_i} C_{d_i}(t)}{2m_i}v_i^2 + \frac{k_{m_i}}{m_i}\right)$$
$$-\frac{\sigma A_i C_{di}(t) v_i a_i}{m_i} \quad (2)$$
$$g_i(v_i) = \frac{1}{m_i \tau_i}$$

where $\tau_i$ is the actuation time-lag; $\sigma$ is the mass of the air; $m_i, A_{F_i}, k_{m_i}, C_{d_i}(t)$ are the mass, cross-sectional area, mechanical drag, and drag coefficient of the vehicle $i$, respectively. By applying the control law in [20], the engine input $\varepsilon_i$ can be derived as (3)

$$\varepsilon_i = u_i m_i + \frac{\sigma A_i C_{d_i}(t)}{2}v_i^2 + k_{m_i} + \tau_i \sigma A_i C_{d_i}(t) v_i a_i \quad (3)$$

where $u_i$ is the control input. After applying the feedback linearization technique [21], $\dot{a}_i(t)$ can be formulated as (4)

$$\dot{a}_i(t) = \frac{1}{\tau_i}(u_i(t) - a_i(t)) \quad (4)$$

By utilizing the vehicle dynamics model, the system state of the leader truck can be defined as $x_0(t) = [d_0(t), v_0(t), a_0(t)]^T$ and the vehicle dynamics can be formulated in a state-space form as (5)

$$\dot{x}_0(t) = A_0 x_0(t) + B_0 u_0(t) + D_0 v_{HDV}(t) \quad (5)$$

where $A_0 = \begin{pmatrix} 0 & -1 & 0 \\ 0 & 0 & 1 \\ 0 & 0 & -1/\tau \end{pmatrix}$, $B_0 = \begin{pmatrix} 0 \\ 0 \\ 1/\tau \end{pmatrix}$, $D_0 = \begin{pmatrix} 1 \\ 0 \\ 0 \end{pmatrix}$ and

$v_{HDV}$ is the velocity of the preceding human-driven vehicle.

The proposed longitudinal control for the leader and the following trucks in the platoon is implemented in a discrete fashion with a zero-order hold (ZOH). The discretized version of (5) with the sample time $T_s$ can be stated as (6)

$$x_0(t+1) = A_0' x_0(t) + B_0' u_0(t) + D_0' v_{HDV}(t) \quad (6)$$

where $A_0'$, $B_0'$ and $D_0'$ can be obtained by applying Jordan-Chevalley decomposition [22].

Since the control objective of each following truck in the platoon is to maintain the constant inter-vehicle distance and zero speed difference to its predecessor, the system error dynamics of each following truck can be derived as (7)

$$\Delta d_i(t) = d_i(t) - d_s$$
$$\Delta v_i(t) = v_{i-1}(t) - v_i(t) \quad (7)$$

where $\Delta d_i$ is the deviation from the equilibrium spacing $d_s$ and $\Delta v_i$ is the speed difference of the vehicle $i$ to its predecessor. The system state of each following truck can be defined as $x_i(t) = [\Delta d_i(t), \Delta v_i(t), a_i(t)]^T$ and the state-space form can be defined as (8)

$$\dot{x}_i(t) = A_i x_i(t) + B_i u_i(t) + D_i a_{i-1}(t) \quad (8)$$

where $A_i = \begin{pmatrix} 0 & 1 & 0 \\ 0 & 0 & -1 \\ 0 & 0 & -1/\tau \end{pmatrix}$, $B_i = \begin{pmatrix} 0 \\ 0 \\ 1/\tau \end{pmatrix}$, $D_i = \begin{pmatrix} 0 \\ 1 \\ 0 \end{pmatrix}$.

The discrete version of (8) with ZOH approach can be defined as (9)

$$x_i(t+1) = A_i' x_i(t) + B_i' u_i(t) + D_i' a_{i-1}(t) \quad (9)$$

*B. Fuel Consumption Model*

In this work, an instantaneous power-based fuel consumption model for heavy-duty trucks [23] is used. This fuel-consumption model has the advantages of easy calibration, high fuel consumption estimation accuracy, and light computation. The fuel-consumption model is formulated as (10)

$$F_{c_i}(t) = \begin{cases} \psi_0 + \psi_1 P_i(t) + \psi_2 P_i(t)^2, & \forall P_i(t) \geq 0 \\ \psi_0, & \forall P_i(t) < 0 \end{cases} \quad (10)$$

where $F_{c_i}$ and $P_i$ are the instantaneous fuel consumption (L/s) and power of the vehicle $i$ at time $t$, respectively, and $\psi_0$, $\psi_1$ and $\psi_2$ are model calibration parameters. The power of vehicle $i$ can be derived as (11)

$$R_i(t) = \frac{\rho}{25.92} C_{d_i}(t) C_a A_{F_i} v_i(t)^2 + 9.8066 m_i \frac{\lambda_0}{1000} \lambda_1 v_i(t)$$
$$+ 9.8066 m_i \frac{\lambda_0}{1000} \lambda_2 + 9.8066 m_i G(t) \quad (11)$$

$$P_i(t) = \left( \frac{R_i(t) + 1.04 m_i a_i(t)}{3600\eta} \right) v_i(t)$$

where $R_i$ is the resistance force, $\rho$ is the air density, $C_a$ is height correction, $\lambda_0$, $\lambda_1$ and $\lambda_2$ are rolling resistance parameters, $G$ is roadway grade, and $\eta$ is vehicle driveline efficiency.

To capture the airflow alteration with variable spacing in the platoon, the drag coefficient $C_{d_i}(t)$ is considered as a nonlinear function with respect to the inter-vehicle distance $d_i(t)$, as evaluated in [24]. This nonlinear relationship is formulated as (12)

$$C_{d_i}(t) = C_{n_i}(t) \left( \gamma_{1_i} d_i(t)^{\gamma_{2_i}} + \gamma_{3_i} \right) \quad (12)$$

where $C_{n_i}$ is the nominal drag coefficient, $\gamma_{1_i}$, $\gamma_{2_i}$ and $\gamma_{3_i}$ are empirical model calibration parameters.

By utilizing this nonlinear relationship, the drag coefficient of each vehicle in the platoon is stated as (13)

$$C_{d_i}(t) = \begin{cases} C_{n_i}, & \text{if } i = 0 \\ C_{n_i} \left( \gamma_{1_i} d_i(t)^{\gamma_{2_i}} + \gamma_{3_i} \right), & \text{else} \end{cases} \quad (13)$$

The vehicle and fuel consumption model parameters are listed in Table I.

### III. CONTROL DESIGN

In this section, the proposed fuel-economical distributed model predictive control (Eco-DMPC) strategy for the heavy-duty truck platoon will be introduced. The control design consists of an eco-driving strategy for the leader truck and a distributed formation control strategy for the following trucks in the platoon.

Table I: The vehicle and fuel consumption model parameters.

| Parameter | Value | Parameter | Value |
|---|---|---|---|
| $m_i$ | 29400 kg | $G$ | 0 |
| $\rho$ | 1.2256 kg/m³ | $\eta$ | 0.94 |
| $C_a$ | 0.977 | $\psi_0$ | 1.56e-3 |
| $C_{n_i}$ | 0.570 | $\psi_1$ | 8.10e-5 |
| $A_{F_i}$ | 10.7 m² | $\psi_3$ | 1.00e-8 |
| $\gamma_{1_1}$ | 0.1522 | $\gamma_{1_{2,3}}$ | 0.0726 |
| $\gamma_{2_1}$ | 0.2111 | $\gamma_{2_{2,3}}$ | 0.2842 |
| $\gamma_{3_1}$ | 0.5260 | $\gamma_{3_{2,3}}$ | 0.5794 |

*A. Eco-Driving Control Design*

The eco-driving strategy is based on a nonlinear model predictive control (NMPC) design integrated with a power-based fuel consumption cost function. This proposed design provides fuel-economical vehicle motion planning for the leader truck by utilizing the speed preview information of the preceding human-driven vehicle, where the speed preview information is assumed to be accessible either via V2V connectivity or driver behavior prediction [13].

The control objective of the eco-driving design is to compute the fuel-economical control input $u_0$ subject to the state and input constraints. The minimum and maximum constraints are applied on the inter-vehicle distance $d_0$ considering the safety and collision avoidance, and allowable distance for vehicle connectivity, respectively. The constraints on the vehicle speed $v_0$ are applied by the determined minimum and maximum speed based on the traffic conditions. At last, the constraints on the vehicle acceleration $a_0$ and control input $u_0$ are imposed for the vehicle's drivability. To solve the fuel-optimal control problem of the leader truck within the short preview horizon at the time $k$, the cost function is formulated as (14)

$$J_0 = \sum_{s=1}^{T_P} F_{c_0}(v_{0,k+s}^{P,k}, a_{0,k+s}^{P,k})$$

subject to:

$$d_{0_{\min}} \leq d_{0,k+s} \leq d_{0_{\max}}$$
$$v_{0_{\min}} \leq v_{0,k+s} \leq v_{0_{\max}} \quad (14)$$
$$a_{0_{\min}} \leq a_{0,k+s} \leq a_{0_{\max}}$$
$$u_{0_{\min}} \leq u_{0,k+s} \leq u_{0_{\max}}$$

where $J_0$ is the accumulated fuel consumption over the prediction horizon $T_P$. At each time step $k$, the proposed NMPC design solves the fuel optimal control problem with the sequential quadratic programming algorithm (SQP) and only the first value of the predicted control input vector $u_{0,k}^P = \left[ u_{0,k}^{P,k}, u_{0,k+1}^{P,k}, ..., u_{0,k+T_P-1}^{P,k} \right]$ is applied to update the vehicle state at the time step $k+1$. Besides, at the time $k$, the predicted state of the leader truck $x_{0,k}^P = \left[ x_{0,k}^{P,k}, x_{0,k+1}^{P,k}, ..., \right.$

$x_{0,k+T_P}^{P,k}$] can be accessible to the following truck 1 under PF communication topology.

### B. Distributed Control Design

In this section, a serial distributed model predictive control (DMPC) for the following trucks under PF communication topology will be introduced. In the proposed serial DMPC design, the following trucks solve the optimal control problem sequentially with their local MPC and each following truck solves the local optimal control by utilizing the predicted information from its leading vehicle through V2V communication. In this study, we assume no communication delay for the sake of simplicity, but such issues can be addressed with existing studies [11], [25].

The following objectives are considered for designing the DMPC strategy for the following trucks in the platoon [26]:

1. All the following trucks in the platoon should be recursive feasible.
2. The entire system should be asymptotically stable. This indicates that all the following trucks with the distributed controller will reach the equilibrium point $\left(x_{i,e}=[0,0,0]^T\right)$ over time when the acceleration of the leader truck is zero.
3. The entire system should be string stable. This refers that the peak magnitude of the spacing error should not be amplified through the vehicular string such as $\|\Delta d_{i+1}\|_\infty \leq \|\Delta d_i\|_\infty$ where $\|\Delta d_i\|_\infty$ defines the $L^\infty$ norm of the spacing error of the following truck $i$.

To solve the optimal control problem of each following truck in the platoon at time $k$, the cost function is formulated as (15)

$$J_i = \sum_{s=1}^{T_P}\left[\left(x_{i,k+s}^{P,k}\right)^T Q_i x_{i,k+s}^{P,k} + W_i\left(u_{i,k+s-1}^{P,k}\right)^2\right] + \left(x_{i,k+T_P}^{P,k}\right)^T Q_{P_i} x_{i,k+T_P}^{P,k}$$

(15)

where $x_{i,k+s}^{P,k}$ is the $(k+s)^{th}$ the predicted state-space, $u_{i,k+s-1}^{P,k}$ is the predicted control input and $x_{i,k+T_P}^{P,k}$ is the predicted terminal state-space of the system at time $k$. $Q_i$ is positive definite weight matrix $Q_i = \begin{pmatrix} \beta_1 & 0 & 0 \\ 0 & \beta_2 & 0 \\ 0 & 0 & \beta_3 \end{pmatrix}$, $W_i$ is positive definite weight value and $Q_{P_i}$ is the terminal cost weight which is the solution of the discrete algebraic Riccati equation to ensure the asymptotic stability of the system, as shown in (16).

$$Q_{P_i} = Q_i + A_i'\left[Q_{P_i} - Q_{P_i}B_i'\left(Q_{P_i} + B_i'^T Q_{P_i} B_i'\right)^{-1} B_i'^T Q_{P_i}\right]A_i' \quad (16)$$

The following control input and state constraints of (15) are defined as follows:

$$\begin{aligned}
u_{i_{\min}} &\leq u_{i,k+s} \leq u_{i_{\max}} & \forall s \in \{0,1,2,...,T_P-1\} \\
a_{i_{\min}} &\leq a_{i,k+s} \leq a_{i_{\max}} & \forall s \in \{1,2,...,T_P\} \\
\Delta d_{i,k+1}^- &\leq \Delta d_{i,k+s} \leq \Delta d_{i,k+1}^+ & \forall s \in \{1,2,...,T_P\}
\end{aligned}$$
(17)

$$\Delta d_{i,k+1}^- = \begin{cases} -d_m & \text{for } i=1 \\ -\max\left(\left|\Delta d_{i-1,s}\right|\right) & \text{for } i>1 \end{cases} \text{ for } \forall s \in \{0,1,2,k+1\}$$

$$\Delta d_{i,k+1}^+ = \max\left(\left|\Delta d_{i-1,s}\right|\right) \quad \text{for } i>1 \text{ for } \forall s \in \{0,1,2,k+1\}$$
(18)

where $d_m$ is a predefined spacing error, the constraint of the spacing error guarantees the $L^\infty$ stability which will be proved later in the following section. The constraints on the control input and acceleration are imposed in the same manner as defined in (14). The proposed DMPC design uses the implementation strategy as introduced in Algorithm 1.

---

**Algorithm 1: The DMPC algorithm**

**Input:** The leader truck's predicted longitudinal acceleration
**Output:** The following trucks' trajectories
1: At time $k=0$,
2:   Initialize $x_{i,0}=0$ for all the following trucks
3: At time $k>0$
4:   **for** $i=1$
5:     The following truck $i$ receives the predicted longitudinal acceleration of the leader $a_{0,k}^P$ and the following truck $i$ derives the predicted vehicle state $x_{i,k}^P$ by solving (15)
6:     Following truck $i$ transmits $a_{i,k}^P$ to $i+1$
7:   **end**
8:   **for** $1<i\leq 3$
9:     The following truck $i$ receives the predicted longitudinal acceleration $a_{i-1,k}^P$ from the following truck $i-1$ and the following truck $i$ derives the predicted vehicle state $x_{i,k}^P$ by solving (15)
10:    **if** $i<3$
11:      Following truck $i$ transmits $a_{i,k}^P$ to $i+1$
12:    **end**
13:  **end**
14: Update time $k=k+1$ and go to step 3

---

### C. Platoon Stability

In the previous section, we formulated the DMPC as a control strategy for the following trucks in the platoon. Next, we will derive the feasibility and the stability of the proposed DMPC design with the following lemmas and proofs.

**Lemma 1:** The DMPC algorithm for the following trucks is recursive feasible if the initial state is feasible and the following states remain in the feasible set.

**Proof:** Assume that DMPC is initially feasible with respect to the constraints and there exists an optimal sequence for all following trucks $\left[u_{i,0}^{P,0}, u_{i,1}^{P,0}, ..., u_{i,T_p}^{P,0}\right] \in U_i$ with the predicted state $\left[x_{i,1}^{P,0}, x_{i,2}^{P,0}, ..., x_{i,T_p}^{P,0}\right] \in x_{i,0}$ and $x_{i,T_p} \in x_{i,f}$ where $x_{i,f}$ is the terminal state domain. Next time interval, the solution is also feasible for the following trucks with the optimal sequence $\left[u_{i,0}^{P,0}, u_{i,1}^{P,0}, ..., K_f(x_{i,T_p}^{P,0}, a_{i-1,T_p}^{P,0})\right] \in U_i$; $\left[x_{i,2}^{P,0}, x_{i,3}^{P,0}, ..., A_i'x_{i,T_p}^{P,0} + B_i'K_f(x_{i,T_p}^{P,0}, a_{i-1,T_p}^{P,0}) + D_i'a_{i-1,T_p}^{P,0}\right] \in x_{i,1}$ and $A_i'x_{i,T_p}^{P,0} + B_i'K_f(x_{i,T_p}^{P,0}, a_{i-1,T_p}^{P,0}) + D_i'a_{i-1,T_p}^{P,0} \in x_{i,f}$ where $K_f$ is the implicit control law. Therefore, the DMPC algorithm is feasible at all time steps.

**Lemma 2:** The DMPC algorithm is asymptotically stable if the following truck 1 is asymptotically stable and the DMPC algorithm for the following trucks is recursive feasible.

**Proof:** To achieve asymptotic stability for the following truck 1 in the platoon, the optimal cost for the following truck 1 $J_1^*$ is a Lyapunov function satisfying $J_{1,1}^* - J_{1,0}^* < 0$ which indicates that optimal cost is decreasing over time and the terminal cost is assumed as a Lyapunov function inside $x_{i,f}$. At time interval 1, there is a sub-optimal control sequence $\left[u_{1,0}^{P,0}, u_{1,1}^{P,0}, ..., K_f(x_{1,T_p}^{P,0}, a_{0,T_p}^{P,0})\right] \in U_1$ with the predicted trajectory $\left[x_{1,2}^{P,0}, x_{1,3}^{P,0}, ..., A_1'x_{1,T_p}^{P,0} + B_1'K_f(x_{1,T_p}^{P,0}, a_{0,T_p}^{P,0}) + D_1'a_{0,T_p}^{P,0}\right] \in x_1$ and the cost $J_{1,1}$ where $J_{1,1} \geq J_{1,1}^*$ and $J_{1,1} - J_{1,0}^*$ can be expressed as follows:

$$J_{1,1} - J_{1,0}^* =$$
$$\begin{pmatrix} A_1'x_{1,T_p}^{P,0} + B_1'K_f(x_{1,T_p}^{P,0}, a_{0,T_p}^{P,0}) \\ + D_1'a_{0,T_p}^{P,0} \end{pmatrix}^T Q_{P_1} \begin{pmatrix} A_1'x_{1,T_p}^{P,0} + B_1'K_f(x_{1,T_p}^{P,0}, a_{0,T_p}^{P,0}) \\ + D_1'a_{0,T_p}^{P,k} \end{pmatrix}$$
$$+ \left(K_f(x_{1,T_p}^{P,0}, a_{0,T_p}^{P,0})\right)^T W_1 \left(K_f(x_{1,T_p}^{P,0}, a_{0,T_p}^{P,0})\right) + \left(x_{1,T_p}^{P,0}\right)^T Q_1 x_{1,T_p}^{P,0}$$
$$- \left(x_{1,1}^{P,0}\right)^T Q_1 x_{1,1}^{P,0} - u_{1,0}^{P,0}W_1 u_{1,0}^{P,0} - \left(x_{1,T_p}^{P,0}\right)^T Q_{P_1} x_{1,T_p}^{P,0} < 0$$

(19)

Since $Q_1$ and $W_1$ are positive definite, we can rewrite (19) as follows:

$$\begin{pmatrix} A_1'x_{1,T_p}^{P,0} + B_1'K_f(x_{1,T_p}^{P,0}, a_{0,T_p}^{P,0}) \\ + D_1'a_{0,T_p}^{P,0} \end{pmatrix}^T Q_{P_1} \begin{pmatrix} A_1'x_{1,T_p}^{P,0} + B_1'K_f(x_{1,T_p}^{P,0}, a_{0,T_p}^{P,0}) \\ + D_1'a_{0,T_p}^{P,k} \end{pmatrix}$$
$$+ \left(K_f(x_{1,T_p}^{P,0}, a_{0,T_p}^{P,0})\right)^T W_1 \left(K_f(x_{1,T_p}^{P,0}, a_{0,T_p}^{P,0})\right) + \left(x_{1,T_p}^{P,0}\right)^T Q_1 x_{1,T_p}^{P,0}$$
$$- \left(x_{1,1}^{P,0}\right)^T Q_1 x_{1,1}^{P,0} < 0$$

(20)

The final expression is $J_{1,1}^* - J_{1,0}^* < J_{1,1} - J_{1,0}^* < -\left(x_{1,T_p}^{P,0}\right)^T Q_1 x_{1,T_p}^{P,0} - u_{1,0}^{P,0}W_1 u_{1,0}^{P,0} < 0$. Therefore, the optimal cost is decreasing over time, and consequently, the asymptotically local stability is satisfied for the following truck 1 in the platoon. When the following truck 1 in the platoon is asymptotically stable, it will stay at a constant speed when it reaches the equilibrium state. Thus, the second following truck in the platoon will not get any disturbance from the first following truck, and the second following truck can be treated as the first following truck without any disturbance. When the second following truck reaches the equilibrium state, there exists a feasible solution for the second following truck. Thus, the asymptotical stability for the remaining following trucks in the platoon can be satisfied by using induction. More details about the asymptotical stability analysis can be found in [26].

**Lemma 3:** The DMPC algorithm for the following trucks is $L^\infty$ string stable when the DMPC algorithm is recursive feasible.

**Proof:** When the DMPC algorithm is recursive feasible, the state constraint $\Delta d_{i,k+1}^+ = \max\left(\left|\Delta d_{i-1,s}\right|\right)$, $i > 1$ for $\forall s \in \{0,1,2,k+1\}$ is satisfied at all time steps. Therefore, the $L^\infty$ string stability $\left\|\Delta d_{i+1}\right\|_\infty \leq \left\|\Delta d_i\right\|_\infty$ is satisfied when $k \to \infty$.

IV. RESULTS AND DISCUSSION

In this section, the fuel economy and stability performances of the proposed truck platooning control strategy will be investigated. Besides, a comparison study with formation-controlled and human-operated platoons will be conducted to evaluate the fuel economy and road utilization performance of the proposed design. For the sake of conciseness, the following trucks are abbreviated as "FT1", "FT2", and "FT3", respectively, in the results. The Eco-DMPC design parameters are listed in Table II.

*A. Baseline Methods*

We compare our proposed Eco-DMPC with two baseline methods to emphasize the improved performance of fuel economy and road utilization for the truck platoon. The first baseline method is a pure DMPC formation control without the eco-driving strategy, where the previously developed DMPC algorithm is adopted for each truck to maintain the constant gap distance and zero speed difference to its predecessor in the platoon. The same design parameters for the following trucks in the proposed Eco-DMPC design are applied to each truck in the platoon for the DMPC baseline design, as shown in Table II.

The second baseline is a human-driven platoon, where the human-operated trucks are modeled by the Intelligent Driver Model (IDM) [27], a well-established model to simulate microscopic traffic flow. The model is briefly introduced as follows:

$$\dot{v}_i = a_i\left(1 - \left(\frac{v_i}{v_{s_i}}\right)^\delta - \left(\frac{s^*(v_i, \Delta v_i)}{s_i}\right)^2\right) \quad (21)$$

$$s^*(v_i, \Delta v_i) = s_0 + Tv_i + \frac{v_i \Delta v_i}{2\sqrt{a_i b_i}} \quad (22)$$

where $v_i$ is vehicle speed; $a_i$ is the maximum acceleration; $b_i$ is the maximum deceleration; $\delta$ is the acceleration component; $v_{s_i}$ is the cruising speed; $\Delta v_i$ is the speed difference to the preceding vehicle; $s^*$ is the desired inter-vehicle distance; $s_0$ is the jam distance; $T$ is the desired time headway. To replicate the human driving preferences in the platoon, the IDM model parameters for heavy-duty trucks are used, as derived in [28]. The maximum acceleration $a$ and deceleration $b$ are set to 1.14 m/s² and 2.29 m/s² respectively; the desired time headway $T$ is set to 2 s; the jam distance $s_0$ is set to 13.6 m; the cruising speed $v_s$ is set to the maximum speed of the HDV along the trip; the acceleration component $\delta$ is set to 4.

Table II: Eco-DMPC design parameters.

| Parameter | Value | Parameter | Value |
|---|---|---|---|
| $T_s$ | 1 s | $v_{0_{min}}$ | 0 m/s |
| $T_P$ | 10 s | $v_{0_{max}}$ | 36 m/s |
| $u_{i_{min}}$ | -4 m/s² | $d_{0_{min}}$ | 5 m |
| $u_{i_{max}}$ | 4 m/s² | $d_{0_{max}}$ | 45 m |
| $a_{i_{min}}$ | -3 m/s² | $d_s$ | 5 m |
| $a_{i_{max}}$ | 3 m/s² | $d_m$ | 3 m |
| $W_i$ | 2 | $\beta_{1,2,3}$ | 1 |
| $x_{i,0}$ | $[0,0,0]^T$ | $x_{i,f}$ | $[0,0,0]^T$ |

### B. Fuel Economy and Road Utilization Analysis

In this study, the EPA US06 driving cycle [29] is used to define the speed profile for the HDV to simulate the preceding traffic in front of the truck platoon. The truck platoon follows the preceding traffic according to the control objectives as previously introduced in (14) and (15). By utilizing the proposed Eco-DMPC design, we first evaluate the fuel consumption performance of the platoon with different desired inter-platoon gap distances. According to the results in Fig. 2, it is observed that the total fuel economy of the platoon improves when the platoon has smaller desired inter-platoon gap distances. This finding conforms with empirical data that the total fuel consumption of the platoon reduces with smaller inter-platoon gap distances due to reduced drag force acting on the trucks.

We then evaluate the fuel economy performance of the proposed Eco-DMPC design with the DMPC and IDM baselines. Fig. 3 shows the speed, acceleration, and gap distance profiles for the leader truck in the Eco-DMPC, DMPC, and IDM strategies. The results show that the leader truck with the proposed Eco-DMPC design performs with 17.97% better fuel economy compared to the DMPC baseline, as shown in Table III and Table IV. Moreover, each following truck and the overall platoon with the Eco-DMPC design show significantly improved fuel economy compared to the DMPC baseline. The fundamental reasons are that the leader truck with Eco-DMPC design better plans its speed by avoiding unnecessary braking and acceleration maneuvers and utilizes the gap distance to the preceding transient traffic as a cushion to omit strong speed fluctuations to itself and as well as to the following trucks in the platoon. However, in the DMPC baseline, the transient traffic conditions cause the leader truck to have excessive braking and acceleration maneuvers since the control objective of the DMPC baseline is to maintain the constant gap distance and zero speed difference to the preceding transient traffic. Consequently, the overall truck platoon with DMPC design is not able to achieve significantly reduced fuel consumption under the transient traffic conditions even though it has better drag force reduction with a smaller average inter-platoon gap distance compared to the proposed Eco-DMPC design, as shown in Table V.

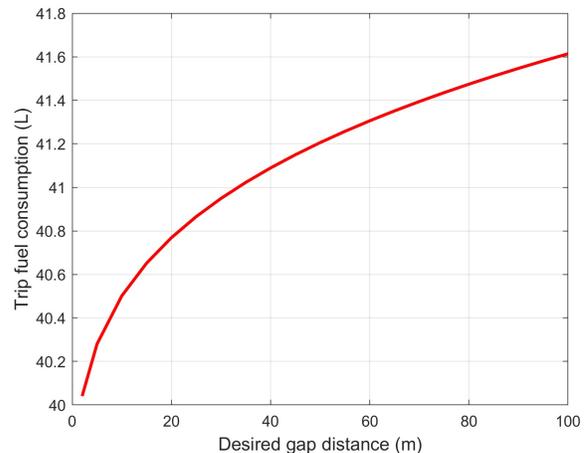

Fig. 2 The variation between the inter-platoon desired gap distance and total fuel consumption of the truck platoon in Eco-DMPC design.

Moreover, the results show that the leader truck with the Eco-DMPC approach better balances its gap distance to the preceding vehicle with a smoother speed profile to minimize fuel consumption by utilizing the preview information compared to the IDM baseline. According to Table IV, the leader truck performs with 9.08% better fuel economy compared to the human-driven leader truck with the IDM baseline. Besides, the Eco-DMPC design performs with approximately 12% to 14% fuel economy improvement compared to the IDM baseline for the following trucks in the platoon. The prime reason for this is two-fold. First, the human-driven truck platoon is not able to efficiently plan vehicle motions without preview information of the highly transient traffic conditions, and consequently, the speed fluctuations during the transient conditions cause more fuel consumption. Second, the truck drivers understandably adopt a larger gap distance when following the preceding vehicle, as shown in Table V, and this preferred larger gap between trucks does not fully benefit from the reduced drag force acting on the trucks. On the other hand, the leader truck with the proposed Eco-DMPC strategy can achieve better fuel-optimal vehicle operations by utilizing the preview information of the transient traffic conditions. Furthermore, the computed small gap distance with the proposed Eco-DMPC design for the following trucks provides efficient drag reduction on the trucks, and the reduced fuel consumption can be achieved accordingly.

Further investigating, the results reveal that truck platoon with IDM baseline performs with better fuel economy compared to the DMPC baseline. The main reason is that the truck platoon modeled with IDM has preferred smoother braking and acceleration maneuvers under transient traffic conditions, thus leading to better fuel economy. In contrast, the truck platoon with the DMPC performs with sharp acceleration and hard braking while achieving its control objectives under transient traffic conditions, which eventually leads to more fuel consumption.

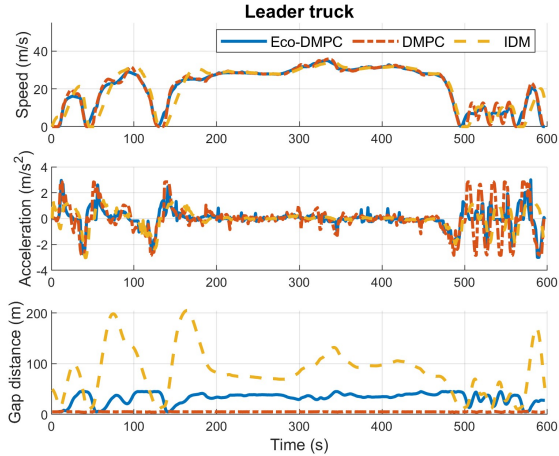

Fig. 3 The comparison of speed, acceleration, and gap distance profiles for the leader truck in Eco-DMPC, DMPC and IDM strategies.

Table III: Trip fuel consumption (L) comparison.

|  | Leader | FT1 | FT2 | FT3 | Total |
|---|---|---|---|---|---|
| Eco-DMPC | 10.47 | 10.06 | 9.88 | 9.87 | 40.28 |
| DMPC | 12.76 | 11.88 | 11.94 | 12.22 | 48.80 |
| IDM | 11.51 | 11.46 | 11.45 | 11.46 | 45.88 |

Table IV: Trip fuel consumption (%) improvement.

| Truck | Eco-DMPC and DMPC (%) | Eco-DMPC and IDM (%) | DMPC and IDM (%) |
|---|---|---|---|
| Leader | 17.97 | 9.08 | -9.79 |
| FT1 | 15.29 | 12.20 | -3.52 |
| FT2 | 17.24 | 13.71 | -4.10 |
| FT3 | 19.25 | 13.88 | -6.23 |
| Total | 17.46 | 12.21 | -5.98 |

Table V: Platoon average gap distance comparison.

|  | Platoon average gap distance (m) |
|---|---|
| Eco-DMPC | 10.56 |
| DMPC | 5.00 |
| IDM | 77.71 |

Another important performance analysis for the proposed control design is road utilization. According to the results in Table V, the truck platoon with the proposed Eco-DMPC design has significantly better road utilization compared to the IDM baseline. The fundamental reason is that the truck drivers prefer a larger gap distance when following the preceding vehicle. In contrast, the proposed Eco-DMPC design provides smaller inter-platoon gap distances by utilizing both eco-driving and platooning strategies. Moreover, the results show that the truck platoon with the DMPC baseline has better road utilization compared to the proposed Eco-DMPC design and the IDM baseline since the DMPC baseline strategy aims to control the platoon with constant and smaller inter-platoon gap distances during the trip.

To summarize the discussion in fuel economy and road utilization analysis, the results show that the proposed Eco-DMPC design can significantly improve the fuel efficiency of the heavy-truck platoon compared to the DMPC and IDM baseline strategies and enhance the road utilization compared to the IDM baseline. The results significantly demonstrate that the proposed Eco-DMPC has the advantages of both eco-driving and platooning strategies to greatly improve the fuel economy and road utilization.

### C. Stability Analysis

In the previous section, we analyzed the fuel economy performance of the proposed Eco-DMPC design. Another important aspect for the performance analysis of the proposed control design is the local and string stability of the truck platoon. Fig. 4 shows the evolution of speed and spacing errors and acceleration of the following trucks during the trip. The results show that the spacing and speed errors and acceleration of the following trucks converge to zero over time when the acceleration of the leader truck is zero, thus satisfying the asymptotical stability requirement. Besides, the $L^\infty$ norm of each following truck is found as 2.29 m, 0.75 m, and 0.56 m, respectively. This means that the truck platoon is satisfying the $L^\infty$ string stability by successfully utilizing the $L^\infty$ norm constraints in (17) **Error! Reference source not found.**, and the errors decay throughout the platoon in the transient traffic conditions.

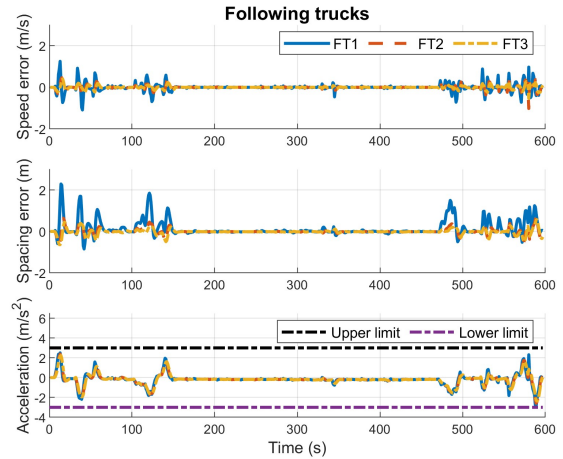

Fig. 4 The speed and spacing errors and acceleration of the following trucks in the platoon with Eco-DMPC design.

## V. CONCLUSION AND FUTURE WORK

In this study, we designed a fuel-efficient distributed model predictive control design for a homogeneous heavy-duty truck platoon with guaranteed local and string stability. The proposed control framework consists of an eco-driving strategy for the leader truck and a distributed control strategy for the following trucks in the platoon. With the eco-driving strategy, the leader truck utilizes the preview traffic conditions to achieve reduced fuel consumption by avoiding energy-inefficient maneuvers during the trip. In the distributed control design, the following trucks maintain the short inter-platoon gap distances to achieve reduced air drag force by ensuring local and string stability. For the comparison study, we evaluated the fuel economy and road utilization performance of the proposed control strategy for the truck platoon with a formation-controlled platoon and a human-driven truck platoon. The comparison results reveal that the heavy-duty truck platoon with the proposed control strategy can achieve significant fuel economy improvement over the formation control strategy and the human baseline model. The results collectively demonstrate that the proposed fuel-efficient distributed control strategy has the potential to improve the energy efficiency and road utilization of freight transportation by synthesizing both eco-driving and platooning strategies.

Indeed, the proposed control strategy is implemented with a nominal control fashion where no disturbance and communication delay in the system are considered. As a further extension of the work, we will design a dedicated control strategy for the truck platoon to explicitly deal with the disturbance and communication delay under different traffic conditions.